\documentclass[12pt]{spieman}  
\usepackage{amsmath,amsfonts,amssymb}
\usepackage{graphicx}
\usepackage{setspace}
\usepackage{lineno}

\title{STROBE-X High Energy Modular Array (HEMA)}

\author[a,*]{Anthony~L.~Hutcheson}
\author[b,c]{Marco~Feroci}
\author[l]{Andrea~Argan}
\author[d]{Matias~Antonelli}
\author[n]{Marco~Barbera}
\author[h]{Jorg~Bayer}
\author[o]{Pierluigi Bellutti}
\author[r]{Giuseppe Bertuccio}
\author[d]{Valter~Bonvicini}
\author[k]{Franck~Cadoux}
\author[p]{Riccardo~Campana}
\author[o]{Matteo~Centis~Vignali}
\author[b]{Francesco~Ceraudo}
\author[a]{Marc~Christophersen}
\author[d]{Daniela Cirrincione}
\author[n]{Fabio D'Anca}
\author[b,k]{Nicolas~De~Angelis}
\author[b]{Alessandra~De~Rosa}
\author[b]{Giovanni~Della~Casa}
\author[b,c]{Ettore~Del~Monte}
\author[b]{Giuseppe~Dilillo}
\author[b,c]{Yuri~Evangelista}
\author[k]{Yannick~Favre}
\author[o]{Francesco~Ficorella}
\author[q]{Mauro~Fiorini}
\author[e]{Jeremy~J.~Ford}
\author[s]{Marco Grassi}
\author[a]{J.~Eric~Grove}
\author[h]{Alejandro~Guzman}
\author[h]{Paul Heddermann}
\author[k,t]{Merlin~R.~Kole}
\author[n]{Ugo~Lo~Cicero}
\author[b]{Giovanni~Lombardi}
\author[s]{Piero Malcovati}
\author[f]{Malgorzata~Michalska}
\author[g]{Aline~Meuris}
\author[l]{Gabriele~Minervini}
\author[f]{Witold~Nowosielski}
\author[b]{Alessio~Nuti}
\author[b]{Luigi~Pacciani}
\author[o]{Giancarlo Pepponi}
\author[e]{Steven~C.~Persyn}
\author[o]{Antonino~Picciotto}
\author[h]{Samuel~Pliego}
\author[d]{Alexander~Rachevski}
\author[d]{Irina~Rashevskaya}
\author[a]{Paul~S.~Ray}
\author[o]{Alina~Samusenko}
\author[h]{Andrea~Santangelo}
\author[g]{Stephane~Schanne}
\author[e]{Carl~L.~Schwendeman}
\author[a]{Clio Sleator}
\author[i]{Jacob~R.~Smith}
\author[j]{Libor~Sveda}
\author[j]{Jiri~Svoboda}
\author[h]{Christoph~Tenzer}
\author[n]{Michela~Todaro}
\author[m]{Alessio~Trois}
\author[d]{Andrea~Vacchi}
\author[h]{Hao~Xiong}
\author[h]{Xianqi~Wang}
\author[k]{Xin~Wu}
\author[a]{Eric~A.~Wulf}
\author[d]{Gianluigi~Zampa}
\author[d]{Nicola~Zampa}
\author[u]{Andrzej~Zdziarski}
\author[o]{Nicola~Zorzi}
\affil[a]{Space Science Division, U.S. Naval Research Laboratory, Washington, DC, USA}
\affil[b]{INAF Institute for Space Astrophysics and Planetology, Rome, Italy}
\affil[c]{National Institute for Nuclear Physics (INFN), Rome Tor Vergata Section, Rome, Italy}
\affil[d]{National Institute for Nuclear Physics (INFN), Trieste Section, Trieste, Italy}
\affil[e]{Southwest Research Institute, San Antonio, TX, USA}
\affil[f]{Space Research Center, Polish Academy of Sciences, Warsaw, Poland}
\affil[g]{CEA Paris-Saclay, DRF/IRFU, Gif-sur-Yvette, France}
\affil[h]{Institute for Astronomy and Astrophysics, Eberhard Karl University of Tübingen, Tübingen, Germany}
\affil[i]{George Mason University, Fairfax, VA, USA}
\affil[j]{Astronomical Institute, Czech Academy of Sciences, Prague, Czech Republic}
\affil[k]{Department of Nuclear and Particle Physics, University of Geneva, Geneva, Switzerland}
\affil[l]{INAF Headquarters, Rome, Italy}
\affil[m]{INAF Astronomical Observatory of Cagliari, Italy}
\affil[n]{INAF Astronomical Observatory of Palermo, Italy}
\affil[o]{Fondazione Bruno Kessler, Povo, Trento, Italy}
\affil[p]{INAF Observatory for Astrophysics and Space Science, Bologna, Italy}
\affil[q]{INAF IASF Milano, Italy}
\affil[r]{Polytechnic of Milan, Italy}
\affil[s]{University of Pavia, Italy}
\affil[t]{University of New Hampshire, Durham, NH, United States}
\affil[u]{Nicolaus Copernicus Astronomical Center, Polish Academy of Sciences, Warszaw, Poland}

\begin{document} 
\maketitle

\begin{abstract}
The High Energy Modular Array (HEMA) is one of three instruments that compose the STROBE-X mission concept. The HEMA is a large-area, high-throughput non-imaging pointed instrument based on the Large Area Detector developed as part of the LOFT mission concept. It is designed for spectral timing measurements of a broad range of sources and provides a transformative increase in sensitivity to X-rays in the energy range of 2--30 keV compared to previous instruments, with an effective area of 3.4 m$^{2}$ at 8.5 keV and an energy resolution of better than 300 eV at 6 keV in its nominal field of regard.
\end{abstract}


{\noindent \footnotesize\textbf{*}\linkable{anthony.l.hutcheson.civ@us.navy.mil} }

\begin{spacing}{1}   

\section{Introduction}
\label{sect:intro}  
\begin{figure}
\begin{center}
\begin{tabular}{c}
\includegraphics[height=10cm]{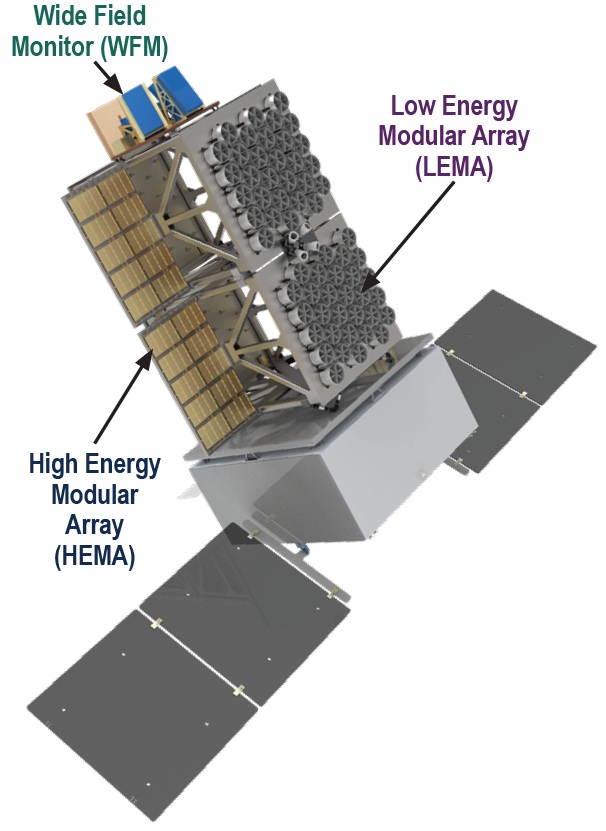}
\end{tabular}
\end{center}
\caption 
{ \label{fig:strobex}
Current STROBE-X design, with the three scientific instruments identified. } 
\end{figure}

The Spectroscopic Time-Resolving Observatory for Broadband Energy X-rays (STROBE-X) \cite{JATISOverview,ray2019strobe} is a probe-class mission concept for time domain and multi-messenger (TDAMM) astrophysics and high-throughput spectroscopy. The goal of this design is broad energy band spectroscopy of time-variable targets and discovery of new sources. STROBE-X comprises three main instruments, as shown in Fig. \ref{fig:strobex}. The High Energy Modular Array (HEMA), the subject of this paper, is a large-area, collimated (non-imaging) instrument operating in the 2--30 keV energy range, complementing the Low Energy Modular Array (LEMA)\cite{JATISLEMA} by providing huge collecting area in the Fe-K line region and above, with an effective area of 3.4 m$^{2}$ at 8.5 keV, approximately 5.5 times greater than that of the Rossi X-ray Timing Explorer (RXTE) Proportional Counter Array (PCA)\cite{Glasser_RXTE_PCA}. The 1$^{\circ}$ field of view (FoV) of the HEMA co-aligns with the LEMA pointing direction. The instrument is based on the technologies of large-area silicon drift detectors (SDDs) and glass micropore collimators, enabling several square meters to be deployed in space within reasonable mass, volume, and power budgets.

\section{Instrument Overview}
\label{sect:overview}  

\begin{figure}
\begin{center}
\begin{tabular}{c}
\includegraphics[height=8cm]{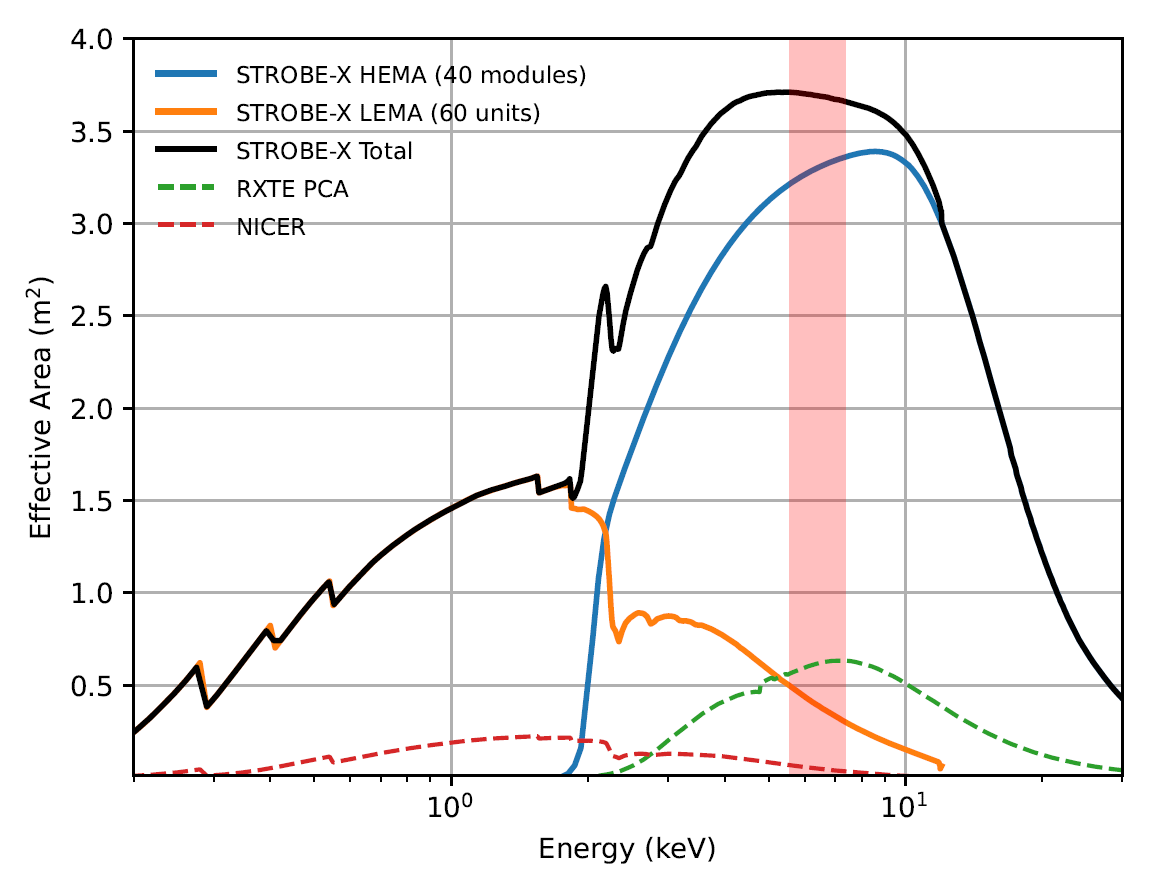}
\end{tabular}
\end{center}
\caption 
{ \label{fig:eff_area}
Effective areas of the STROBE-X HEMA and LEMA compared against the RXTE PCA and NICER, respectively. } 
\end{figure}

The concept and design of the HEMA is based on the Large Area Detector (LAD) developed as part of the  Large Observatory For x-ray Timing (LOFT) \cite{10.1117/12.2054654,2012ExA....34..415F} mission concept and further matured  for the enhanced X-ray Timing and Polarimetry (eXTP) \cite{2022SPIE12181E..1XF,ZhangeXTP} mission concept. As the name implies, the HEMA is highly modular in design, providing the instrument with inherent resilience (no single point of failure), parallelism in manufacture/testing/integration, and smooth scalability (e.g., increasing or decreasing the number of detector modules in the instrument). The instrument is composed of four independent panels -- two on each side of the LEMA -- that each comprise ten modules. Each module houses 16 SDDs, for a total of 640 detectors. This design provides an effective area 5.5 times greater than the RXTE PCA (see Fig. \ref{fig:eff_area}) with approximately 3 times better energy resolution. The modularity and the collimated FoV of the instrument -- 1$^{\circ}$ full width at half maximum (FWHM) -- results in a significant reduction in dead time and pile-up events, making the HEMA an extraordinary instrument for spectral-timing information on bright Galactic sources. Table \ref{tab:hema_char} summarizes the main characteristics of the HEMA. This instrument's capabilities are nicely complemented by the lower-energy sensitivity of the co-aligned LEMA instrument and the imaging capabilities of the Wide Field Monitor (WFM)\cite{JATISWFM}, the sum of these parts resulting in a game-changing observatory for TDAMM astrophysics and high-throughput timing and spectroscopy.

\section{Instrument Design}
\label{sect:design}  
\begin{figure}
\begin{center}
\begin{tabular}{c}
\includegraphics[height=5.5cm]{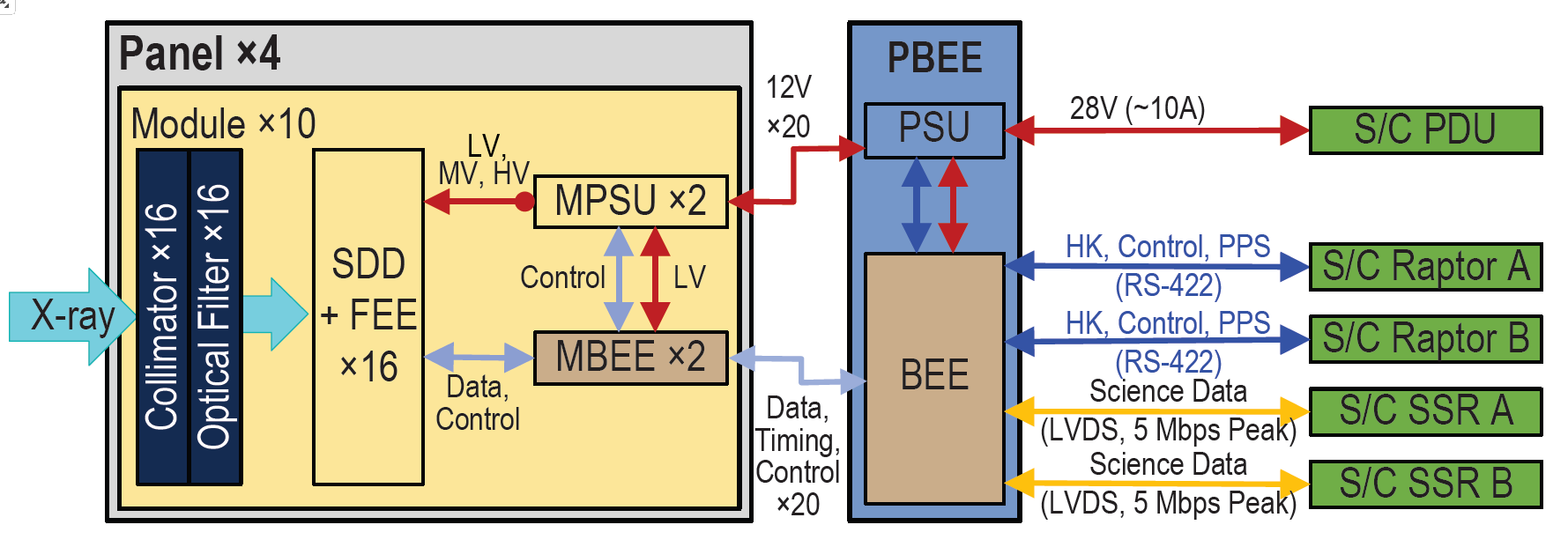}
\end{tabular}
\end{center}
\caption 
{ \label{fig:func_block}
Functional block diagram of the HEMA instrument.} 
\end{figure}

The hierarchical design of the HEMA instrument is shown in Fig. \ref{fig:func_block}. Incident X-ray are collimated and pass through an optical blocking filter before impinging on the SDD. Data flow from each SDD to an associated front-end electronics (FEE) board, where events are digitized before passing to the module back-end electronics (MBEE). Two MBEEs each serve eight SDD/FEE assemblies in a given module, time stamping and packaging the events to transmit to the panel back-end electronics (PBEE) and, ultimately, to the spacecraft. Instrument power flows from the spacecraft to a power supply unit at the panel level, where it is redistributed to each of the ten modules in that panel. Module power supply units (MPSUs) are responsible for converting the 12-V input from the panel supply into the appropriate low, medium, and high voltages required by the SDDs and readout electronics in that module. In the following subsections, the individual components of the HEMA will be described, starting the detector level and moving up in hierarchy.

\subsection{HEMA Modules}

\subsubsection{Detectors}

X-ray detection for the HEMA is performed with the use of large area SDDs\cite{Rachevski_2014}. These monolithic SSDs (active area: 108.5 mm $\times$ 70.0 mm, geometric area: 120.8 mm $\times$ 72.5 mm, thickness: 0.45 mm) are being jointly developed in Italy by the National Institute for Nuclear Physics (INFN), Fondazione Bruno Kessler (FBK), the National Institute for Astrophysics (INAF), and the Italian Space Agency (ASI). The nominal energy range for these detectors is 2--30 keV, though, for bright sources outside the nominal FOV (sufficient to overcome both the X-ray attenuation of the collimator walls and the lower detector efficiency at higher X-ray energies), the range could be extended up to 80 keV. Furthermore, the good energy resolution (300--500 eV at 5.9 keV) and time resolution (10 $\mu$s) of these detectors make them excellent choices for X-ray astrophysics in the target energy range.

\begin{figure}
\begin{center}
\begin{tabular}{c}
\includegraphics[height=6.6cm]{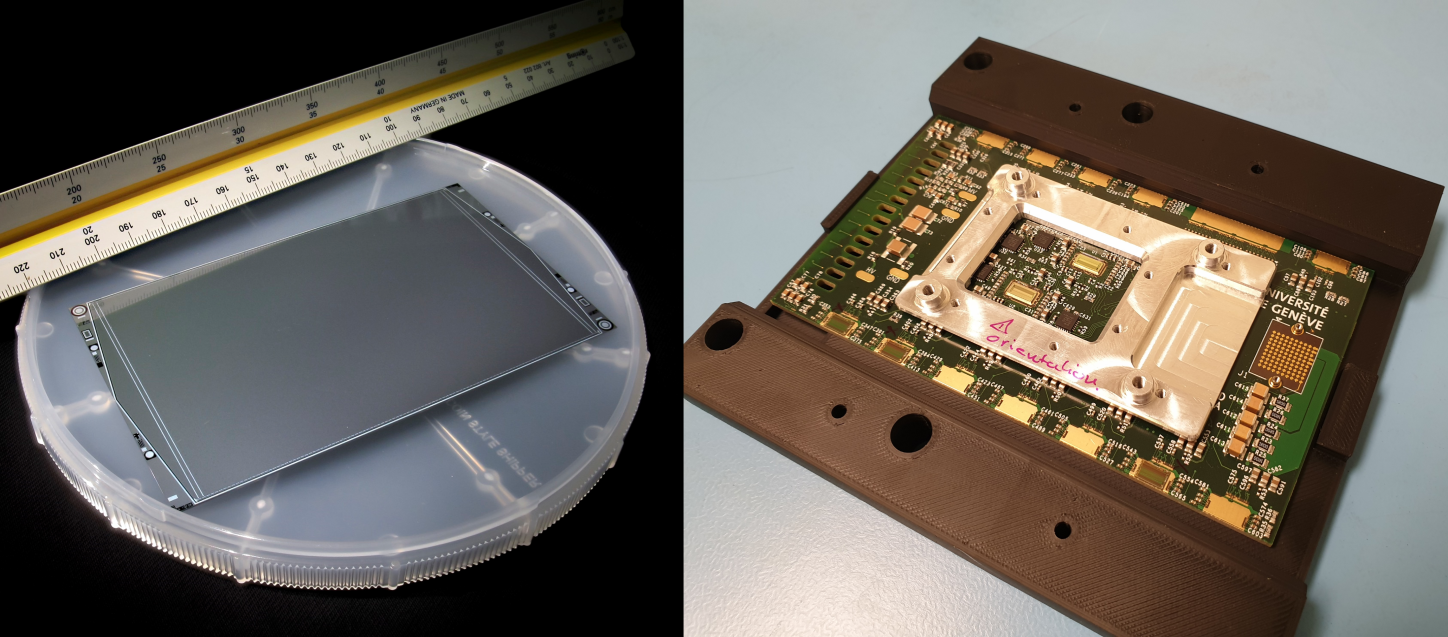}
\\
(a) \hspace{8cm} (b)
\end{tabular}
\end{center}
\caption 
{ \label{fig:sdd_fee}
(a) HEMA silicon drift detector. (b) HEMA front-end electronics board. } 
\end{figure}

Figure \ref{fig:sdd_fee}(a) shows an SDD manufactured by FBK. X-rays that interact with the detector create charge clouds that are drifted to the long edges of the detector by an electric field generated by cathodes biased to 1300 V. The arrangement of cathodes divides the detector into halves, so that the maximum drift length for a given charge to reach a detector edge is 35 mm. The charge at each edge is read out by 112 anodes on a 970 $\mu$m pitch, for a total of 224 anodes per detector. Full-scale prototypes of this detector have been manufactured by FBK and tested extensively by INFN and INAF. For the STROBE-X HEMA, the SDDs are further designed to confine the drifting charges so that $>$95\% of events trigger only a single anode, with an input capacitance of 90 fF. FBK produced full-scale detectors (geometric area: 120.3 mm $\times$ 72.4 mm) with this updated design and delivered them to INFN for characterization in February 2024.

\subsubsection{Front-end electronics}

The HEMA FEE, developed by the University of Geneva, is responsible for readout and analog-to-digital conversion of the charge signal from the SDD anodes and hosts the necessary bias connectors for the SDD. Each HEMA SDD has an associated FEE board onto which it is mounted; the FEE board is glued directly to the SDD, additionally providing mechanical support to the SDD. Figure \ref{fig:sdd_fee}(b) shows one such board. Each FEE board houses 16 application-specific integrated circuits (ASICS), eight ASICs for each detector edge that is populated by 112 anodes; thus, each ASIC is responsible for reading out 14 SDD anodes via wire bond connections. Due to the large pitch of the SDD anodes (970 $\mu$m), the number of anodes per ASIC was limited to 14 to avoid a large ASIC width. These ASICs are being developed at the U.S. Naval Research Laboratory (NRL) and leverage design heritage from previous front-end ASIC designs \cite{10338415}. Details of the ASIC design are presented in Ref.~\citenum{JATISASIC}. Each ASIC channel comprises a dual-stage low-noise charge amplifier, a shaper with baseline stabilizer, an event discriminator with trimmable threshold, a peak detector, control logic, local configuration registers and a multiplexer. Within an ASIC, the channels share bias circuits, digital-to-analog converters (DACs) for test pulsers and threshold generators, common configuration registers and a temperature sensor. The power dissipation is 0.590 mW per channel, with ~6 mW dissipation from the common circuitry. In parallel, the French Alternative Energies and Atomic Energy Commission (CEA) is designing an alternative ASIC based on the IDeF-X HD design \cite{GEVIN2012415} to mitigate development risk as the NRL ASICs currently have a low technical readiness level (TRL) compared to the other HEMA components; having a backup ASIC option reduces the technical risk associated with this component. Because the count rate per channel is expected to be sufficiently low, an analog-to-digital converter (ADC) is not integrated into the ASIC; instead, two external ADCs on the FEE board can multiplexed to serve eight ASICs, each ADC therefore serving half of the detector.

\subsubsection{Collimators}

\begin{figure}
\begin{center}
\begin{tabular}{c}
\includegraphics[height=5.8cm]{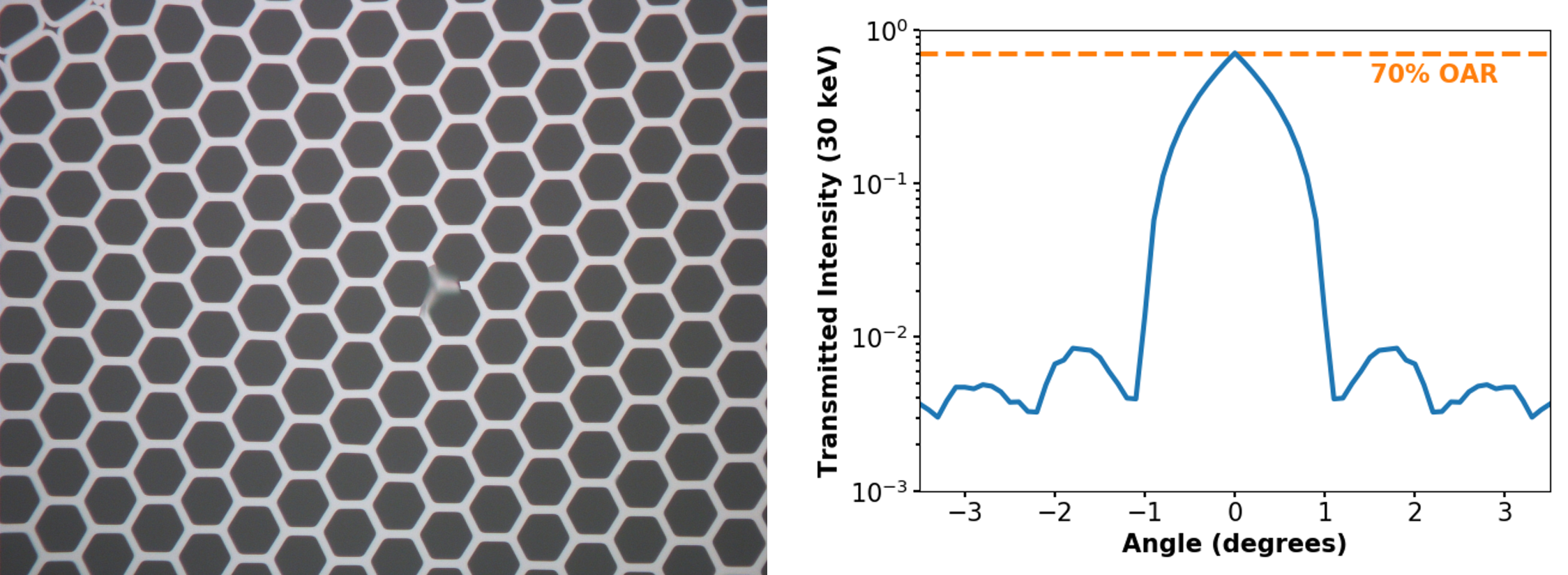}
\\
(a) \hspace{8cm} (b)
\end{tabular}
\end{center}
\caption 
{ \label{fig:collimator}
(a) Optical micrograph of Incom glass capillary collimator. (b) Simulated angular response of collimator against 30-keV X-rays. } 
\end{figure}

Each HEMA detector is paired with a glass micropore collimator of equal area that limits the field of view (FOV) to 1$^{\circ}$ FWHM. Produced by Incom, Inc. (Charlton, MA, USA), the collimator is a 5-mm-thick micropore plate composed of leaded glass. Figure \ref{fig:collimator}(a) shows an optical micrograph of one of these micropore plates. The hexagonal pores have an apothem of 41.5 $\mu$m and a pitch of 99 $\mu$m, resulting in an open area ratio of 70\%. Figure \ref{fig:collimator}(b) shows the simulated angular response of the collimator against 30-keV X-rays\cite{9032903} . At the upper limit of the nominal HEMA energy range, the leaded glass of the collimator attenuates $>$99\% of incident X-rays outside of the desired FOV.

\subsubsection{Optical filters}
In addition to X-rays, the HEMA SDDs are sensitive to photons in the optical regime, which increases dark current and thus readout noise. To achieve the low noise needed to effectively detect X-rays, the SDDs must be shielded from optical photons. The requirement to block this light from reaching the SDDs is primarily achieved by the mechanical structure of the HEMA module, but in the pointing direction of the HEMA instrument, this requirement must be balanced with the need for X-ray transparency. For this purpose, HEMA uses optical blocking filters manufactured by Luxel Corporation (Friday Harbor, WA, USA). The filters comprise 200 nm of aluminum deposited on self-standing (no support mesh) 2-$\mu$m polyimide films and provide the required attenuation in the optical band without degrading the X-ray performance ($>$87.5\% transmission at 2.5 keV). The filter design has been tested extensively \cite{cicero2022filters} for this purpose, and testing of sample filters from Luxel is planned for the near future.

\subsubsection{Module back-end electronics}
\begin{figure}
\begin{center}
\begin{tabular}{c}
\includegraphics[height=5.5cm]{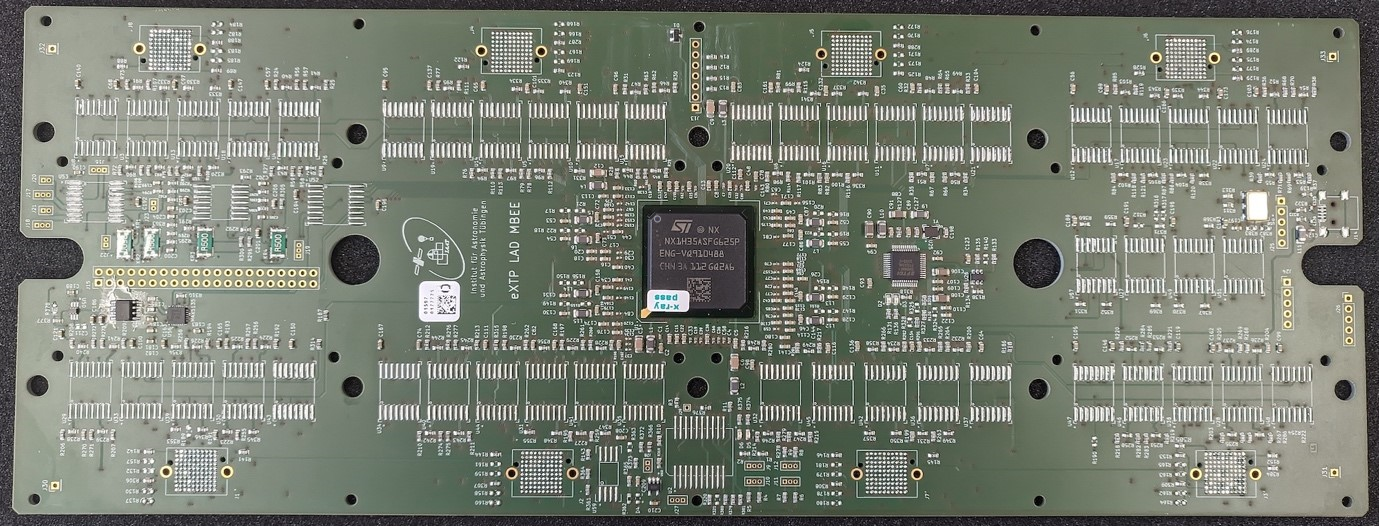}
\end{tabular}
\end{center}
\caption 
{ \label{fig:hema_mbee}
Photo of a HEMA MBEE demonstrator board. } 
\end{figure}

Each module houses two MBEEs, each serving one half of the module (eight SDDs and FEEs). Figure \ref{fig:hema_mbee} shows a MBEE demonstrator board fabricated by IAAT University of T{\"u}bingen. This board includes a radiation-hardened field programmable gate array (FPGA), low-voltage differential signaling (LVDS) transceivers, interfaces to other relevant HEMA components (FEEs, PBEE, and MPSU), and voltage and current monitor circuits. The MBEE is responsible for collecting digitized events from, and distributing telecommands to, the FEEs. Digitized events are pre-processed \cite{xiong2022digital}, time stamped, and packaged for transmission to the PBEE. In addition, the MBEE is responsible for acquiring housekeeping data for transmission to the PBEE as well as control and distribution of power to the FEEs.

\subsubsection{Module structure}

\begin{figure}
\begin{center}
\begin{tabular}{c}
\includegraphics[height=18cm]{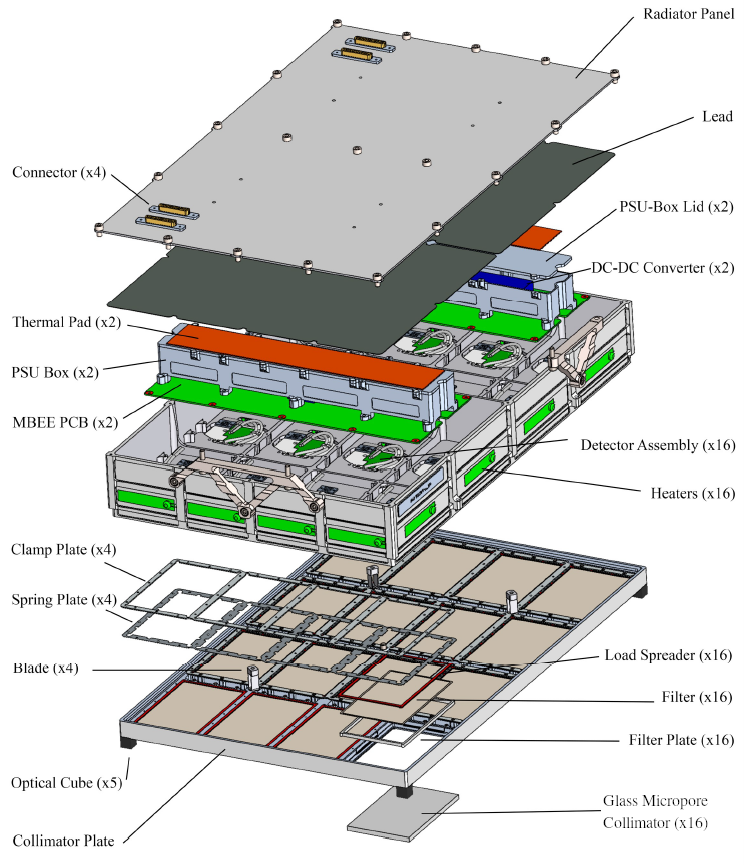}
\end{tabular}
\end{center}
\caption 
{ \label{fig:hema_module}
Exploded view of the HEMA module design. } 
\end{figure}

Figure \ref{fig:hema_module} shows an exploded view of the HEMA module, which hosts the previously described components and serves as the core functional element of the HEMA. The module structure itself is divided into two parts: a collimator tray and a detector tray. The former is responsible for providing mechanical support for, and maintaining co-alignment of, the 16 collimators and optical filters of the module; co-alignment to within 1 arcmin is maintained by a clamping mechanism designed to accommodate the different thermal expansion coefficients of the glass micropore collimator and the aluminum alloy frame of the tray. The detector tray provides mechanical support for the 16 SDD/FEE assemblies of the module as well as the associated MBEEs and MPSUs. The rear of the module comprises an aluminum alloy radiator panel to dissipate heat and a 300-$\mu$m lead shield to reduce background events in the detectors. Thermal pads facilitate heat transfer from the MPSUs toward the rear radiator panel. Each module is thermally isolated from panel support frame. The HEMA modules have no active cooling but do incorporate heaters to maintain the temperature above the low operational limit ($-55$ $^{\circ}$C); these heaters are also designed to allow the SDDs to be annealed periodically (see Sec. \ref{sect:radtherm}). Optical cubes are attached to the collimator frame to allow alignment of the modules during integration into the panel structure. 

\subsection{HEMA Panels}

\subsubsection{Panel back-end electronics}

\begin{figure}
\begin{center}
\begin{tabular}{c}
\includegraphics[height=5.5cm]{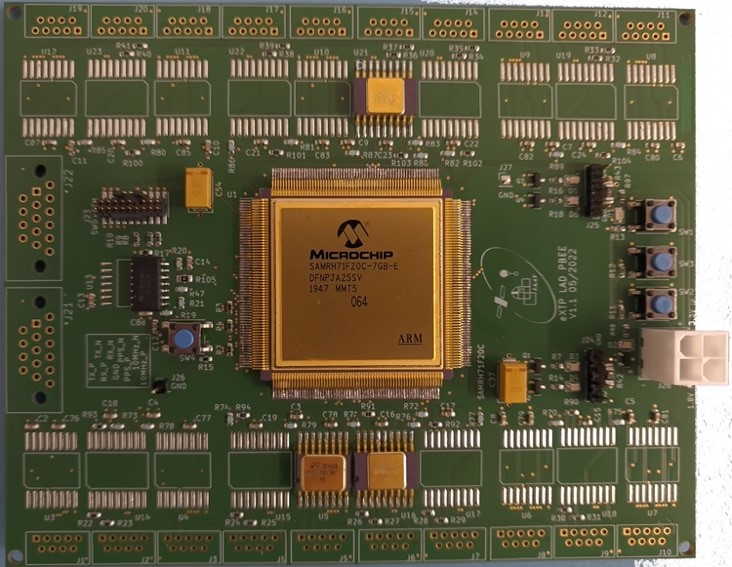}
\end{tabular}
\end{center}
\caption 
{ \label{fig:hema_pbee}
Photo of a HEMA PBEE demonstrator board. } 
\end{figure}

Each panel hosts a single PBEE that handles all events from the ten modules in its panel and provides the power and data interface directly to the STROBE-X spacecraft. Figure \ref{fig:hema_pbee} shows a PBEE demonstrator board fabricated by IAAT University of T{\"u}bingen. Each PBEE has redundant RS422 universal asynchronous receiver-transmitter (UART) and pulse per second (PPS) interfaces to communicate command and telemetry with the spacecraft avionics as well as a redundant LVDS high speed serial interface to transmit high-rate science data directly to the spacecraft solid state recorder. The PBEE's main tasks are interfacing the MBEE, collecting and buffering event packets, reformatting timestamps and binning data depending on the observation mode, transferring data to the spacecraft, executing spacecraft commands,
and collecting and formatting telemetry housekeeping packets sent to the spacecraft. The standard operational mode for HEMA telemeters full information about each event. This includes event type, differential time tag, and event energy. In parallel, a set of ratemeters will be collected with four energy bands and 16 ms integration time. For very bright sources, the PBEE can switch to a binned mode with reduced time or energy resolution.

\subsubsection{Panel structure}

\begin{figure}
\begin{center}
\begin{tabular}{c}
\includegraphics[height=8cm]{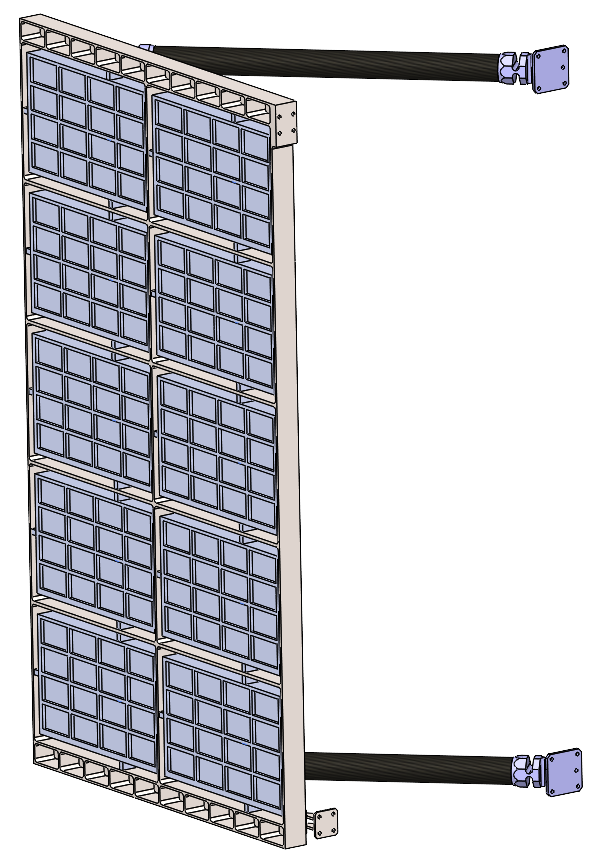}
\end{tabular}
\end{center}
\caption 
{ \label{fig:hema_panel}
Rendering of a complete HEMA panel. } 
\end{figure}

Figure \ref{fig:hema_panel} shows a rendering of a HEMA panel. Each HEMA panel supports ten modules and one PBEE on a fixed aluminum structure that provides the required mechanical alignment for the modules as well as co-alignment with the LEMA boresight to within 3 arcmin. The panels have provisions for harnessing that connect the PBEE to the modules without interfering with the individual module radiators. The monolithic structure allows for structural efficiency to be easily and reliably determined, reducing risk and uncertainty in the design.

\section{Calibration}
\label{sect:calibration}  
INAF will perform energy scale and resolution calibration versus temperature for the HEMA SDDs at the detector tray level by illuminating the full tray with uncollimated X-rays. Collimator tiles and filters will be acceptance tested to verify throughput and angular response at NRL before shipping to INAF. After integrating the complete module (detector tray plus collimator tray), the effective area (versus energy and angle) will be calibrated using the 30-m beamline facility in Palermo.

During on-orbit commissioning, all HEMA channels will be tested and calibrated using injected signals from the front-end ASICs, and alignment will be calibrated using slews across bright point sources. The energy scale (gain and offset) will be calibrated using injected signals and observations of X-ray spectral source targets (e.g., Cas A SNR). Additional long-term gain monitoring will be performed with the 10.5-keV L fluorescence line from the Pb atoms in the glass collimators. During the mission, repeated observations will allow characterization of the variation with temperature and time (from radiation damage). For background, blank fields are chosen (as for RXTE and NICER) and monitored regularly to enable background model fits. The background variation over the orbit is $<$10\%, and modeling of effects such as off-axis source leakage shows that a systematic error should be 0.3\% (against a 1\% requirement), except for a few targets with a very bright source just outside the FoV.

\section{On-Orbit Radiation Environment and Thermal Considerations}
\label{sect:radtherm}  
One of the major contributing factors to the HEMA energy resolution is electronic noise, which has a substantial contribution from the leakage current from the SDD anodes. At beginning of life (BOL), this leakage current is expected to be 680 pA/anode at 20 $^{\circ}$C. To determine the expected leakage current at end of life (EOL), accumulated radiation damage was modeled using AP9\cite{8101531}. For a 15$^{\circ}$ inclination orbit, this damage adds about 4075 pA anode per year (at 20 $^{\circ}$C, without any annealing). In addition, the leakage current is reduced by a factor of 2 for every 7 $^{\circ}$C reduction in detector temperature. At BOL, this means that the resolution is $<$300 eV (FWHM) at 6 keV for a detector equivalent temperature of $-10$ $^{\circ}$C or lower. Without mitigation, by EOL, a temperature of $-35$ $^{\circ}$C would be required to achieve a resolution of $<$500 eV. However, the effects of radiation damage can be mitigated by annealing the SDDs \cite{MOLL2002100}. For HEMA, annealing at +49 $^{\circ}$C for two total days every six months (one day per two panels) would relax the EOL temperature requirement to $-25$ $^{\circ}$C.

\begin{figure}
\begin{center}
\begin{tabular}{c}
\includegraphics[height=8cm]{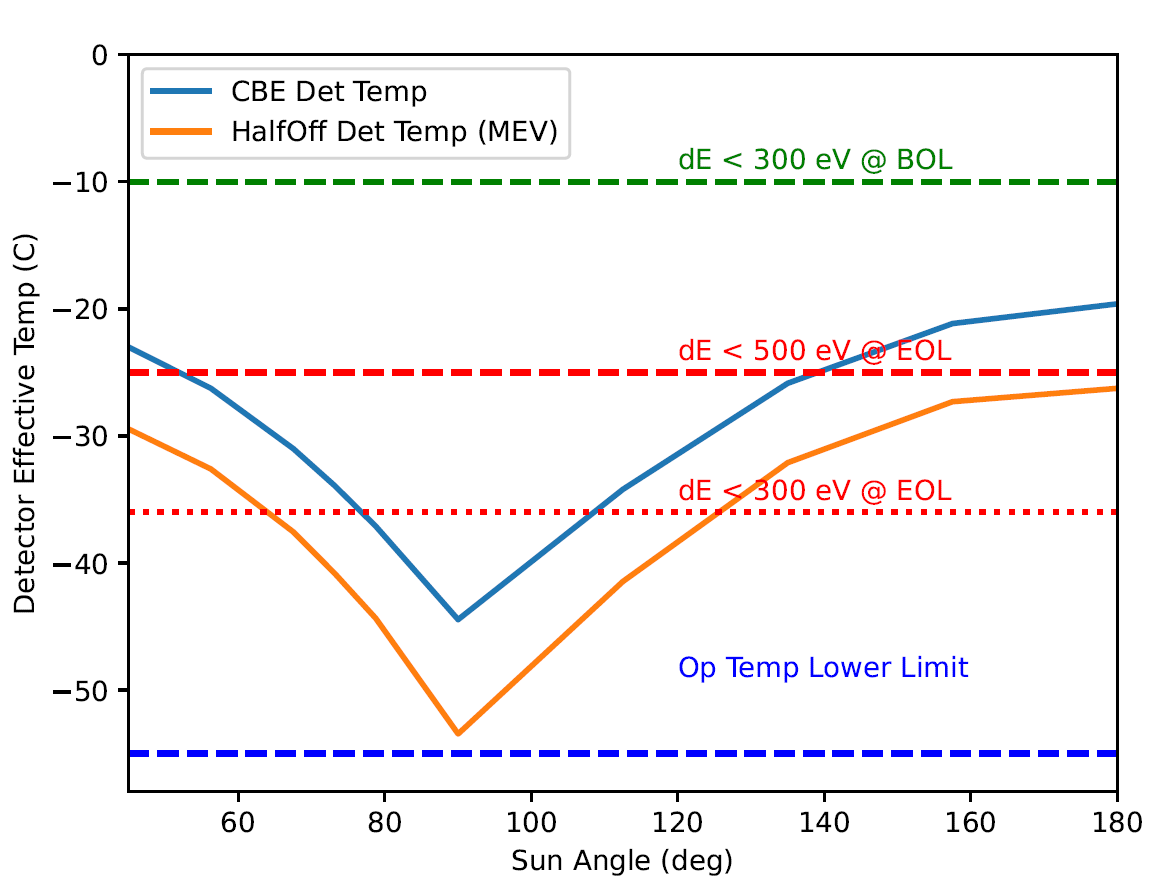}
\end{tabular}
\end{center}
\caption 
{ \label{fig:hema_temp}
HEMA detector temperature as a function of Sun angle for a 15$^{\circ}$ inclination orbit at an altitude of 575 km. Blue is the current best estimate (CBE) of the orbit-average detector temperature with all detectors operating. The orange line shows the maximum temperature expected with half the detector off, extending the sun angle range to the full 45$^\circ$--180$^\circ$ even at end-of-life. Dashed lines indicate temperatures required to achieve different resolution conditions. (Note: the dashed EOL lines assume the HEMA SDDs are annealed every six months).} 
\end{figure}

Figure \ref{fig:hema_temp} shows the detector temperature as a function of Sun angle ($\Sigma$) for an inclination of 15$^{\circ}$ and an altitude of 575 km, with indications of the temperatures required to achieve a resolution at BOL and EOL. We define the field of regard (FoR) as the Sun angle range where the resolution can be maintained at $<$500 eV. At BOL, this is 45$^{\circ}$ $<$ $\Sigma$ $<$ 180$^{\circ}$ (85\% of sky). At EOL, the FoR drops to 60$^{\circ}$ $<$ $\Sigma$ $<$ 140$^{\circ}$ (63\% of sky). We define an extended FoR (eFoR) where a resolution of $<$500 eV can be achieved by turning off half of the HEMA detectors per module, which covers 45$^{\circ}$ $<$ $\Sigma$ $<$ 180$^{\circ}$ at EOL. This option allows observations of high priority, time-critical targets, albeit with reduced HEMA collecting area. It should be noted that, for short observations, the thermal inertia is sufficient such that observing in a cool orientation for some time, then pointing to a source in a hotter orientation, can keep the detectors operating at their goal resolution for ~2 ks. Because the resolution will always be a function of time and temperature, the response matrix generation software will account for this, and the calibration database will have calibration data over a wide range of leakage currents.

\section{Technology Status and Development Plan}
\label{sect:hema_trl}  
The HEMA instrument concept has been in development for a decade, with significant progress made during the LOFT Assessment Phase study and for the eXTP Concept Study. HEMA strongly leverages the technical maturity and development experience associated with this previous work to minimize technical risk. Some changes from previous iterations are required for STROBE-X, where the NRL team will provide the collimators, optical blocking filters, and baseline front-end ASICs. Final layout of the ASIC is complete, and fabrication will be complete by late-2024. U.S. industry leading commercial vendors have been identified for the filters (Luxel) and collimators (Incom) using flight-proven and off-the-shelf manufacturing techniques, with small form factor changes for HEMA. The HEMA module engineering is ongoing under the leadership of INAF, with technology demonstrator models for FEE, MBEE, PBEE, and mechanical structure already in hand and an end-to-end test of the digital electronics chain fully demonstrated. Full-scale prototypes of the HEMA SDD have been produced and are currently undergoing characterization. A full technical, cost, and schedule plan with margin is in place for building and testing a HEMA detector module prior to PDR scheduled for November 2027) to achieve TRL-6 maturity.

\section{Conclusions}
The High Energy Modular Array is a large-area X-ray instrument that contributes to the game-changing capabilities offered the STROBE-X mission concept. Its transformative sensitivity for X-rays in the energy range of 2-30 keV is enabled by over a decade of development for the LOFT and eXTP missions. The instrument concept -- leveraging the recently developed technology of large area SDDs -- provides significant improvements in both effective area and energy resolution over previous instruments while also providing inherent resilience through its modular design. Its enormous collecting area, when combined with co-aligned LEMA instrument, provides unique access to precision measurements of physical conditions in nature’s most extreme environments. The technology is well developed, with all subsystems either already at TRL-6 or with a clear development plan to demonstrate TRL-6 by March 2026.

\subsection*{Disclosures}
The authors have no potential conflicts of interest to disclose.

\subsection* {Code, Data, and Materials Availability} 
Data sharing is not applicable to this article, as no new data were created or analyzed.

\subsection* {Acknowledgments}
Work performed at NRL was sponsored by the Office of Naval Research (ONR) as part of 6.1 funding and by NASA Astrophysics Research and Analysis (APRA) program (22-APRA22-0026);the STROBE-X mission concept study was funded by the NASA Astrophysics Probes program (16-APROBES16-0008). Italian authors acknowledge support by the Italian Space Agency through grant 2020-3-HH.1. J.S. acknowledges PRODEX PEA 4000144379, GACR project 21-06825X and institutional support from RVO:67985815.


\bibliography{report}   
\bibliographystyle{spiejour}   


\vspace{2ex}\noindent\textbf{Anthony Hutcheson} is an research physicist at the U.S. Naval Research Laboratory. He received his BS degree in physics from the University of Georgia in 2000 and his MA and PhD degrees in physics from Duke University in 2002 and 2008, respectively.  His current research interests include novel technologies and techniques to detect ionizing radiation in space.

\vspace{2ex}\noindent\textbf{Marco Feroci} is an astrophysicist and Director at the Institute for Space Astrophysics and Planetology (IAPS) of the Italian National Institute for Astrophysics (INAF). He received his BS degree in physics from the University of Rome Sapienza in 1992. He is author of more than 250 refereed papers on astrophysics of Galactic compact objects, gamma ray bursts and detector technology for X-rays. 

\vspace{1ex}
\noindent Biographies and photographs of the other authors are not available.

\clearpage

\begin{table}[ht]
\caption{Characteristics of the HEMA instrument.} 
\label{tab:hema_char}
\begin{center}       
\begin{tabular}{|l|l|} 
\hline
\rule[-1ex]{0pt}{3.5ex}  \textbf{Instrument Characteristic} & \textbf{Value}  \\
\hline\hline
\rule[-1ex]{0pt}{3.5ex}  Energy Range & 2--30 keV (Nominal) \\
\rule[-1ex]{0pt}{3.5ex}   & 2--80 keV (Extended) \\
\hline
\rule[-1ex]{0pt}{3.5ex}  Effective Area & 3.4 m$^{2}$ at 8.5 keV \\
\hline
\rule[-1ex]{0pt}{3.5ex}  Field of View & 1$^{\circ}$ FWHM \\
\hline
\rule[-1ex]{0pt}{3.5ex}  Field of Regard (FoR) & 45$^{\circ}$ -- 180$^{\circ}$ Sun Angle (BOL) \\
\rule[-1ex]{0pt}{3.5ex}   & 60$^{\circ}$ -- 140$^{\circ}$ Sun Angle (EOL) \\
\hline
\rule[-1ex]{0pt}{3.5ex}  Extended Field of Regard (eFoR) & 45$^{\circ}$ -- 180$^{\circ}$ Sun Angle \\
\hline
\rule[-1ex]{0pt}{3.5ex}  Absolute Time Accuracy (to UTC) & 7 $\mu$s  \\
\hline
\rule[-1ex]{0pt}{3.5ex}  Energy Resolution & $<$300 eV at 6 keV (BOL) \\
\rule[-1ex]{0pt}{3.5ex}   & $<$500 eV at 6 keV (EOL)\\
\hline
\rule[-1ex]{0pt}{3.5ex}  Maximum Source Flux & 15 Crab  \\
\hline
\rule[-1ex]{0pt}{3.5ex}  Maximum Expected Mass & 743.1 kg  \\
\hline
\rule[-1ex]{0pt}{3.5ex}  Maximum Expected Power & 1128.6 W \\
\hline
\rule[-1ex]{0pt}{3.5ex}  Telemetry Rate &  2450 kbps (1 Crab) \\
\rule[-1ex]{0pt}{3.5ex}   &  15 kbps (Background) \\
\hline 
\end{tabular}
\end{center}
\end{table}


\clearpage

\end{spacing}
\end{document}